# Ferromagnetism in Mn-doped $Sb_2Te$

Huixia Luo, Quinn Gibson, Jason Krizan and R. J. Cava

Department of Chemistry, Princeton University, Princeton, New Jersey 08544, USA

**Abstract**

We report that $Sb_2Te$, a natural superlattice phase consisting of two elemental $Sb_2$ layers interleaved with single $Sb_2Te_3$ layers, becomes ferromagnetic at low temperatures on doping with small percentages of Mn. Ferromagnetism appears for Mn concentrations as low as $Sb_{1.98}Mn_{0.02}Te$, where a ferromagnetic $T_c$ of ~ 8.6 K is observed. $T_c$ decreases with increasing Mn content in the stoichiometric materials but increases with increasing Te excess in materials of the type $Sb_{1.93-y}Mn_{0.07}Te_{1+y}$, starting at ~ 3 K at y = 0 and reaching a $T_c$ of ~ 8.9 K at y = 0.06.

## 1. Introduction

Dilute magnetic semiconductors (DMSs), which are prepared by doping transition metals into nonmagnetic semiconductor hosts, have attracted considerable interest for both their scientific value and their potential spintronics applications. Ferromagnetism in semiconductors at low temperatures has been reported for various II-VI [1,2], IV-VI [3,4], and III-V [5,6] compounds doped with transition metals such as Mn, Cr, V, and Fe. Recently, it has been experimentally shown that topological insulators based on the small band gap semiconductors $Bi_2Te_3$ [7,8] and $Sb_2Te_3$ [9,10] become ferromagnetic at low temperatures when doped with Mn, as does Cr-doped $(Bi,Sb)_2(Te,Se)_3$ [11,12]. $Bi_{2-x}Mn_xTe_3$, for example, attains a $T_c$ = 12 K for x = 0.09, and the Curie temperature of $Sb_{1.985}Mn_{0.015}Te_3$ is reported to reach as high as 17 K. $Sb_2Te$ is a small band gap semiconductor in the Sb-Te system that is also expected to display topological surface states [13]; it is a natural superlattice phase made of layers of the elemental semimetal Sb and the normal valence semiconductor $Sb_2Te_3$. Because the interaction of topological surface states with ferromagnetism is expected to yield interesting effects, it is therefore of interest to determine whether Mn doping of $Sb_2Te$, i.e. in $Sb_{2-x}Mn_xTe$, induces ferromagnetism at low temperatures.

First principles electronic structure calculations predict that the parent $Sb_2Te$ material is a small band gap semiconductor and a topological insulator [13,14]. It is one member of the $(Sb_2)_n(Sb_2Te_3)_m$ infinitely adaptive series, whose crystal structure (see inset of Figure 1a) can be regarded as a natural superlattice in the c-axis direction. The structure of $Sb_2Te$ consists of single quintuple layers of $Sb_2Te_3$ stacked with two bi-layer sandwiches of Sb, i.e. in the layer stacking sequence -{[Te-Sb-Te-Sb-Te]-[Sb-Sb]–[Sb-Sb]}-. The $Sb_2$ layers and their stacking are similar to those found in elemental Sb [15-17]. This compound crystallizes in the trigonal space group P3m1 (#164) with a = 4.272(1) Å, c = 17.633(1) Å [18].

Here we report the synthesis and characterization of Mn-doped $Sb_2Te$ in polycrystalline form. Curie temperatures of around 8.6 K are observed at very low doping levels, i.e. for $Sb_{1.98}Mn_{0.02}Te$. For higher Mn contents, we find that the ferromagnetic $T_c$ depends on the Sb:Te ratio; $Sb_{1.89}Mn_{0.07}Te_{1.04}$ and $Sb_{1.87}Mn_{0.07}Te_{1.06}$ display $T_c$s near 7 and 8.9 K, respectively.

## 2. Experimental

Polycrystalline $Sb_{2-x}Mn_xTe$ (x = 0, 0.005. 0.01, 0.02, 0.04, 0.06, 0.07) and $Sb_{1.93-y}Mn_{0.07}Te_{1+y}$ (y = 0, 0.02, 0.04, 0.06) samples were prepared by a solid state reaction method. Stoichiometric quantities of high-purity elemental Sb (99.9999 %), Mn (99.99 %) and Te (99.999 %) were sealed in clean evacuated quartz ampoules. The Mn was purified of oxygen before use. All the ampoules were heated up to 1000 $^{o}$C, where the samples were melted and held overnight, after which they were quenched in water. Silver-colored polycrystalline boules were obtained. The samples were confirmed to be single phase by powder X-ray diffraction (XRD) using a Bruker D8 diffractometer with Cu K$\alpha$ radiation and a graphite diffracted beam monochromator. DC magnetization and Hall measurements were performed in a Quantum Design Physical Property Measurement System (PPMS).

**3. Results and Discussion**

Figures 1b and c show the powder XRD patterns of the polycrystalline $Sb_{2-x}Mn_xTe$ (x = 0, 0.005. 0.01, 0.02, 0.04, 0.06, 0.07) and $Sb_{1.93-y}Mn_{0.07}Te_{1+y}$ (y = 0, 0.02, 0.04, 0.06) samples, respectively. All the Mn-doped $Sb_2Te$ samples studied show single phase $Sb_2Te$ in the diffraction patterns - no impurity MnTe or $MnTe_2$ peaks were observed. For slow cooled rather than quenched samples, Mn-Te phases were present in the X-ray patterns, indicating a lower Mn solubility limit in the $Sb_2Te$ phase at lower temperatures.

Figure 2a shows the zero-field cooled temperature dependence of the magnetic susceptibility for the polycrystalline $Sb_{2-x}Mn_xTe$ (x = 0.005. 0.01, 0.02, 0.04, 0.06, 0.07) series, under an applied magnetic field, H = 10 kOe. The inset of Figure 2a emphasizes the susceptibilities of $Sb_{2-x}Mn_xTe$ at low temperature. The susceptibilities for $Sb_{2-x}Mn_xTe$ (x = 0.005, 0.01, 0.02) are negative at high temperature, due to core diamagnetism. All the Mn-doped crystals have a paramagnetic susceptibility at low temperature, which is ascribed to the increasing paramagnetic contribution from the doped Mn. A transition to a ferromagnetic state is seen in the $Sb_{2-x}Mn_xTe$ samples at low temperatures, as the observed magnetization becomes positive and large below $T_c$. Above the transition temperature, the magnetic susceptibility ($\chi$) can be fit to the Curie-Weiss law, $\chi$-$\chi_0$ = C/(T-$\theta$), where C is the Curie constant, $\theta$ is the paramagnetic Curie temperature, and $\chi_0$ is the temperature independent contribution to the susceptibility. The fits were performed in the temperature range of 40 – 100 K; the M vs. H curves at 40 K and higher are linear to beyond applied fields of 10 kOe (Figure 3a), showing the validity of employing $\chi$ = M/H at H = 10 KOe. The effective magnetic moment ($P_{eff}$) per Mn ion can be obtained by using $P_{eff} = (8C)^{1/2}$. Figure 2b shows the low temperature inverse susceptibility plots, $1/(\chi-\chi_0)$ vs. T. The inverse susceptibilities for x = 0.005 and 0.01 show straight line behavior, which is due to the weak local moment from the small amount of doped-Mn. As the Mn doping increases, the effective moment per mol-Mn decreases. The fitted parameters are summarized in Table 1.

Figure 3b shows the field-dependent magnetization curves at 2 K for $Sb_{2-x}Mn_xTe$ (x = 0.02, 0.04, 0.06, 0.07) samples, with an increasing field from 0 to 90 kOe. The magnetization values reach ~ 0.05, 0.08, 0.088 and 0.09 $\mu_B$ f.u.$^{-1}$ at 2 K at 90 kOe, respectively. A comparison of narrow hysteresis loops on varying the applied field from -3 to 3 kOe for $Sb_{2-x}Mn_xTe$ (x = 0.02, 0.04, 0.06, 0.07) samples is shown in Figure 4. Small coercivities ($H_c$s) of ~ 300 Oe, 380 Oe, 420 Oe and 420 Oe were observed for the $Sb_{1.98}Mn_{0.02}Te$, $Sb_{1.96}Mn_{0.04}Te$, $Sb_{1.94}Mn_{0.06}Te$, and $Sb_{1.93}Mn_{0.07}Te$ samples, respectively, which indicate that the samples are soft ferromagnets. The $H_c$s for our $Sb_{2-x}Mn_xTe$ (x = 0.02, 0.04, 0.06, 0.07) samples are similar to the previously reported single crystal $Bi_{1.91}Mn_{0.09}Te_3$ ($H_c$ ~ 350 Oe, 2 K) [19] samples, although they are much lower than has been reported for V-doped $Sb_2Te_3$ ($H_c$ ~ 12 kOe, 2 K) [20]. The $Sb_2Te$ materials with higher Mn content show greater remnant magnetization, $M_r$. For x = 0.07, a remnant magnetization $M_r$ = 0.0085 $\mu_B$ f.u.$^{-1}$is obtained. For x = 0.06, x = 0.04 and x = 0.02, $M_r$ = 0.0074, $M_r$ = 0.0060 and $M_r$ = 0.0037 $\mu_B$ f.u.$^{-1}$, respectively.

To better determine the temperature of the ferromagnetic transitions, we apply the Arrott plot criterion for $T_c$ determination in disordered systems [21]. Figure 5 presents the Arrott plots for polycrystalline $Sb_{2-x}Mn_xTe$ (x = 0.02, 0.04, 0.06, 0.07) samples. In such plots, H/M versus M should behave as $H/M = A'(T-T_c)+B'M^2+C'M^4$ at values of H above those of the magnetic hysteresis. At $T_c$, the intercept changes sign. For second order transitions, B' is positive. On the other hand, theory predicts that the (high-field) isotherms of $M^2$ vs H/M are parallel lines for ferromagnets and the isotherm at $T_c$ passes through zero. As shown in Figure 5, the ferromagnetic transition in $Sb_{1.98}Mn_{0.02}Te$ is second order as expected and the $T_c$ is close to 8.6 K. The ferromagnetic transition temperatures for the higher Mn-content materials $Sb_{1.94}Mn_{0.06}Te$ and in $Sb_{1.93}Mn_{0.07}Te$ (see Figures 5c, d) are lower, ~ 3 K.

We hypothesized that the $T_c$ for $Sb_{1.98}Mn_{0.02}Te$ is greater than that for $Sb_{1.93}Mn_{0.07}Te$ in spite of its lower Mn content due to the fact that for the higher Mn-content samples the *p*-type carrier concentration is too high. Therefore we designed a second series of materials, $Sb_{1.93-y}Mn_{0.07}Te_{1+y}$ (y = 0, 0.02, 0.04, 0.06), for comparison. In this case, up to a doping level where Fermi energy pinning results in no further change, the *p*-type carrier concentration is expected to decrease as Te is substituted in the Sb site, donating electrons.

Figure 6a shows the zero-field cooled temperature dependence of the magnetic susceptibilities for the polycrystalline $Sb_{1.93-y}Mn_{0.07}Te_{1+y}$ (y = 0, 0.02, 0.04, 0.06) samples with the applied magnetic field, H = 10 kOe. The inset of Figure 6a emphasizes the susceptibilities of polycrystalline $Sb_{1.93-y}Mn_{0.07}Te_{1+y}$ (y = 0, 0.02, 0.04, 0.06) at low temperature. All the $Sb_{1.93-y}Mn_{0.07}Te_{1+y}$ samples have a paramagnetic susceptibility at low temperature, which is consistent with the polycrystalline $Sb_{2-x}Mn_xTe$ samples, and is ascribed to the paramagnetic contribution from the Mn. The comparison of the low temperature inverse susceptibility plots, $1/(\chi-\chi_0)$ vs. T is shown in Figure 6b. The fitted parameters are summarized in Table 2. Again, the fits were performed in the temperature range of 40 – 100 K; the M vs. H curves at 40 K and higher are linear to beyond applied fields of 10 kOe (Figure 7a), showing the validity of the use of $\chi$ = M/H at H = 10 KOe.

The field-dependent magnetization curves at 2 K for polycrystalline $Sb_{1.93-y}Mn_{0.07}Te_{1+y}$ (y = 0, 0.02, 0.04, 0.06) on increasing field from 0 to 90 kOe are shown in Figure 7b. Their magnetizations values reach ~ 0.09, 0.09, 0.08 and 0.1 $\mu_B$ f.u.$^{-1}$ at 2 K at 90 kOe for $Sb_{1.93-y}Mn_{0.07}Te_{1+y}$ (y = 0, 0.02, 0.04, 0.06), respectively. A comparison of the narrow hysteresis loops on varying the applied field from -3 to 3 kOe for polycrystalline $Sb_{1.93-y}Mn_{0.07}Te_{1+y}$ (y = 0, 0.02, 0.04, 0.06) samples is shown in Figure 8. Comparing the four samples, the coercivities ($H_c$s) are the same, but the remnant magnetization $M_r$ changes; the higher the Te content, the larger the remnant magnetization ($M_r$) values.

Figure 9 presents the Arrott plots for the samples of $Sb_{1.93-y}Mn_{0.07}Te_{1+y}$ (y = 0, 0.02, 0.04, 0.06). The data show a systematic increase in $T_c$ with increasing Te content, beginning at ~ 3 K for $Sb_{1.93}Mn_{0.07}Te$ and reaching a maximum of ~ 8.9 K for $Sb_{1.87}Mn_{0.07}Te_{1.06}$. This is consistent with our expectation that excess Te acts as an *n*-type dopant and therefore that the *p*-type carrier concentration for $Sb_{1.93}Mn_{0.07}Te$ is too high to support the maximum $T_c$ possible at that Mn doping content in this phase. This conclusion is confirmed by our measurements of the carrier concentrations of the Te-excess samples. All samples are *p*-type. The carrier concentrations for the $Sb_{1.93-y}Mn_{0.07}Te_{1+y}$ materials are found to be 3.4 x $10^{21}$ $cm^{-3}$, 2.0 x $10^{21}$ $cm^{-3}$, 0.6 x $10^{21}$ $cm^{-3}$, and 1.2 x $10^{21}$ $cm^{-3}$ for y = 0, 0.02, 0.04 and 0.06 respectively. Thus $T_c$ increases as the *p*-type carrier concentration decreases. The Te is not an efficient dopant, only a small number of electrons are produced per number of Te dopants, and increased Te concentration beyond a few percent excess does not decrease the net carrier concentration to low values or to *n*-type behavior. This is consistent with the well-known difficulty in attaining low *p*-type concentrations or *n*-type behavior in the semiconductor $Sb_2Te_3$ [16,22].

## 4. Conclusions

A series of substitution of Mn-doped $Sb_2Te$ has been investigated. A Curie temperature of ~ 8.6 K is observed in $Sb_{1.98}Mn_{0.02}Te$; this is a relatively low magnetic ion content (1 % of the metal sites) for inducing ferromagnetism in a semiconductor. $T_c$ decreases with increasing Mn content in materials of the type $Sb_{2-x}Mn_xTe$ up to the solubility limit of x = 0.07, but higher $T_c$'s are again obtained for x = 0.07 when excess Te is employed to adjust the hole carrier concentration. This ferromagnetic topological insulator system provides an avenue for studying the interactions of ferromagnetism with topological surface states. Further study would be of interest.

## Acknowledgements

This work was supported by the DARPA MESO program grant N66001-11-1-4110.

**References**

1. A. Haury, A. Wasiela, A. Arnoult, J. Cibert, S. Tatarenko, T. Dietl, and Y. Merle d'Aubigné, Phys. Rev. Lett. **79**, 511 (1997).
2. H. Saito, V. Zayets, S. Yamagata, and K. Ando, Phys. Rev. Lett. **90**, 207202 (2003).
3. T. Story, R. R. Gała□zka, R. B. Frankel, and P. A. Wolff, Phys. Rev. Lett. **56**, 777 (1986).
4. Y. D. Park, A. T. Hanbicki, S. C. Erwin, C. S. Hellberg, J. M. Sullivan, J. E. Mattson, T. F. Ambrose, A. Wilson, G. Spanos, and B. T. Jonker, Science **295**, 651 (2002).
5. D. Kitchen, A. Richardella, J. M. Tang, M. E. Flatte, and A. Yazdani**,** Nature **442,** 436 (2006).
6. P. Mahadevan, and A. Zunger, Phys. Rev. B **68,** 075202 (2003).
7. Y. S. Hor, P. Roushan, H. Beidenkopf, J. Seo, D. Qu, J. G. Checkelsky, L. A. Wray, D. Hsieh, Y. Xia, S. Y. Xu, D. Qian, M. Z. Hasan, N. P. Ong, A. Yazdani, and R. J. Cava, Phys. Rev. B **81**, 195203 (2010).
8. L. Cheng, Z. G. Chen, S. Ma, Z. d. Zhang, Y. Wang, H. Y. Xu, L. Yang, G. Han, K. Jack, G. Lu, and J. Zou, J. Am. Chem. Soc. **134**, 18920 (2012).
9. J. S. Dyck, P. Svanda, P Lostak, J Horak, W Chen, and C Uher, J. Appl. Phys. **94**, 7631 (2003).
10. J. Choi, S. Choi, J. Choi, Y. Park, H. M. Park, H. W. Lee, B. C. Woo, and S. Cho, Phys. Status Solidi (b) **241,** 1541 (2004).
11. J. Zhang, C. Z. Chang, P. Tang, Z. Zhang, X. Feng, K. Li, L. L. Wang, X. Chen, C. Liu, W. Duan, K. He, Q. K. Xue, X. Ma, and Y. Wang, Science, **339**, 1582 (2013).
12. C. Z. Chang, J. Zhang, X. Feng, J. Shen, Z. Zhang, M. Guo, K. Li, Y. Ou, P. Wei, L. L. Wang, Z. Q. Ji, Y. Feng, S. Ji, X. Chen, J. Jia, X. Dai, Z. Fang, S. C. Zhang, K. He, Y. Wang, L. Lu, X. C. Ma, and Q. K. Xue, Science **340**, 167 (2013).
13. Q. Gibson et al. to be published.
14. Y. Takagaki, A. Giussani, J. Tominaga, U. Jahn, and R. Calarco, J Phys.: Condens. Matter **25**, 345801 (2013).
15. C. S. Barrett, P. Cucka, and K. Haefner, Acta Cryst. **16,** 451 (1963).
16. R. J. Cava, H. Ji, M. K. Fuccillo, Q. D. Gibson, and Y. S. Hor, J. Mater. Chem. C **1**, 3176 (2013).
17. V. Agafonov, N. Rodier, R. Céolin, R. Bellissent, C. Bergman and J. P. Gaspard, Acta Cryst. **C47**, 1141-1143 (1991).
18. V. Agafonov, N. Rodier, R. Céolin, R. Bellissent, C. Bergman, and J. P. Gaspard, Acta Crystallogr. Sect. C **47**, 1141 (1991).
19. Y. S. Hor, P. Roushan, H. Beidenkopf, J. Seo, D. Qu, J. G. Checkelsky, L. A. Wray, D. Hsieh, Y. Xia, S. Y. Xu, D. Qian, M. Z. Hasan, N. P. Ong, A. Yazdani, and R. J. Cava, Phys. Rev. B **81**, 195203 (2010).
20. J. S. Dyck, P. Hájek, P. Lošt'ák, and C. Uher, Phys. Rev. B **65**, 115212 (2002).
21. I. Yeung, R. M. Roshko, and G. Williams, Phys. Rev. B **34**, 3456 (1986).
22. Z. Starý, J. Horák, M. Stordeur, and M. Stölzer, J. Phys. Chem. Solids **49**, 29 (1988).

**Figures Captions**

**Figure 1.** (Color online) (a) The crystal structure of $Sb_2Te$. The unit cell is shown as solid lines. The Sb and Te atoms are plotted as blue and red spheres, respectively. The $Sb_2$ bi-layer and $Sb_2Te_3$ layer building blocks are described in ball-and-stick and polyhedral forms, respectively; (b) The powder XRD patterns for $Sb_{2-x}Mn_xTe$ (x = 0, 0.005. 0.01, 0.02, 0.04, 0.06, 0.07) and (c) $Sb_{1.93-y}Mn_{0.07}Te_{1+y}$ (y = 0, 0.02, 0.04, 0.06).

**Figure 2.** (Color online) (a) Zero-field cooled temperature-dependent dc magnetic susceptibility measured at 10 kOe applied magnetic field for the $Sb_{2-x}Mn_xTe$ (x = 0.005. 0.01, 0.02, 0.04, 0.06, 0.07) samples. (b) Temperature dependence of the inverse susceptibility for the $Sb_{2-x}Mn_xTe$ (x = 0, 0.005. 0.01, 0.02, 0.04, 0.06, 0.07) samples.

**Figure 3.** (Color online) (a) The magnetic-field-dependent magnetization of the $Sb_{2-x}Mn_xTe$ (x = 0, 0.005. 0.01, 0.02, 0.04, 0.06, 0.07) at 40 K. (b) The magnetic-field-dependent magnetization of $Sb_{2-x}Mn_xTe$ (x = 0.02, 0.04, 0.06, 0.07) at 2 K.

**Figure 4.** (Color online) The hysteresis M vs. H loops of $Sb_{2-x}Mn_xTe$ (x = 0.02, 0.04, 0.06, 0.07) samples at 2 K with applied field varied from -3 to 3 kOe.

**Figure 5.** (Color online) Arrott plots at various temperatures near $T_c$ for the polycrystalline $Sb_{2-x}Mn_xTe$, (a) x = 0.02, (b) x = 0.04, (c) x = 0.06, (d) x = 0.07 samples. (e) Inverse susceptibility (a') inferred from the intercept of Arrott plots ($H/M = a' + b'M^2$) of Figure 5a, b, c, d. (f) Detail of the regime near $T_c$ of Figure 5e.

**Figure 6.** (Color online) (a) Zero-field cooled temperature-dependent dc magnetic susceptibility measured at 10 kOe applied magnetic field for the polycrystalline $Sb_{1.93-y}Mn_{0.07}Te_{1+y}$ (y = 0, 0.02, 0.04, 0.06) samples. (b) Temperature dependence of the inverse susceptibility for the polycrystalline $Sb_{1.93-y}Mn_{0.07}Te_{1+y}$ (y = 0, 0.02, 0.04, 0.06) samples.

**Figure 7.** (Color online) (a) The magnetic-field-dependent magnetization of the polycrystalline $Sb_{1.93-y}Mn_{0.07}Te_{1+y}$ (y = 0, 0.02, 0.04, 0.06) at 40 K. (b) The magnetic-field-dependent magnetization of the polycrystalline $Sb_{1.93-y}Mn_{0.07}Te_{1+y}$ (y = 0, 0.02, 0.04, 0.06) samples at 2 K.

**Figure 8.** (Color online) The hysteresis MH loops of the polycrystalline $Sb_{1.93-y}Mn_{0.07}Te_{1+y}$ (y = 0, 0.02, 0.04, 0.06) samples at 2 K with applied field from -3 to 3 kOe.

**Figure 9.** (Color online) Arrott plots at various temperatures near $T_c$ for the polycrystalline $Sb_{1.93-y}Mn_{0.07}Te_{1+y}$ (a) y = 0, (b) y = 0.02, (c) y = 0.04, (d) y = 0.06 samples. (e) Inverse susceptibility (a') inferred from the intercept of Arrott plots ($H/M = a' + b'M^2$) of Figure 9a, b, c, d. (f) Detail of the regime near $T_c$ of Figure 9e.

**Table 1**. Parameters from fitting the magnetic susceptibility of $Sb_{2-x}Mn_xTe$ (x = 0, 0.005. 0.01, 0.02, 0.04, 0.06, 0.07) to the Curie-Weiss law and Arrott plots.

| x | C (emu $mol^{-1}$ $Oe^{-1}$ $f.u^{-1}$) | $\chi_0$ (emu $mol^{-1}$ $Oe^{-1}$ $f.u.^{-1}$) | θ (K) | $P_{eff}$ ($\mu_B$/mol-Mn) | $T_c$ (K) | M at 90 kOe at 2 K ($\mu_B$/f.u.) |
|---|---|---|---|---|---|---|
| 0.005 | 0.01963 | -0.00042 | -5.05 | 5.60 | - | - |
| 0.01 | 0.03770 | -0.00040 | -0.85 | 5.49 | - | - |
| 0.02 | 0.06639 | -0.00035 | 5.57 | 5.15 | 8.6 | 0.05 |
| 0.04 | 0.10278 | -0.000013 | 3.04 | 4.53 | 4.1 | 0.08 |
| 0.06 | 0.10834 | -0.000215 | 5.29 | 3.8 | 3.6 | 0.088 |
| 0.07 | 0.11587 | 0.000015 | 2.89 | 3.63 | 3 | 0.09 |

**Table 2**. Parameters from fitting the magnetic susceptibility of $Sb_{1.93-y}Mn_{0.07}Te_{1+y}$ (y = 0, 0.02, 0.04, 0.06) to the Curie-Weiss law and Arrott plots.

| y | C (emu $mol^{-1}$ $Oe^{-1}$ $f.u^{-1}$) | $\chi_0$ (emu $mol^{-1}$ $Oe^{-1}$ $f.u.^{-1}$) | θ (K) | $P_{eff}$ ($\mu_B$/mol-Mn) | $T_c$ (K) | M at 90 kOe at 2 K ($\mu_B$/f.u.) |
|---|---|---|---|---|---|---|
| 0 | 0.11587 | 0.000015 | 2.89 | 3.63 | 3 | 0.09 |
| 0.02 | 0.10384 | 0.000059 | 8.19 | 3.44 | 5 | 0.09 |
| 0.04 | 0.10042 | -0.00024 | 8.07 | 3.42 | 7 | 0.08 |
| 0.06 | 0.09985 | 0.000151 | 15.76 | 3.38 | 8.9 | 0.1 |

**Figure 1**

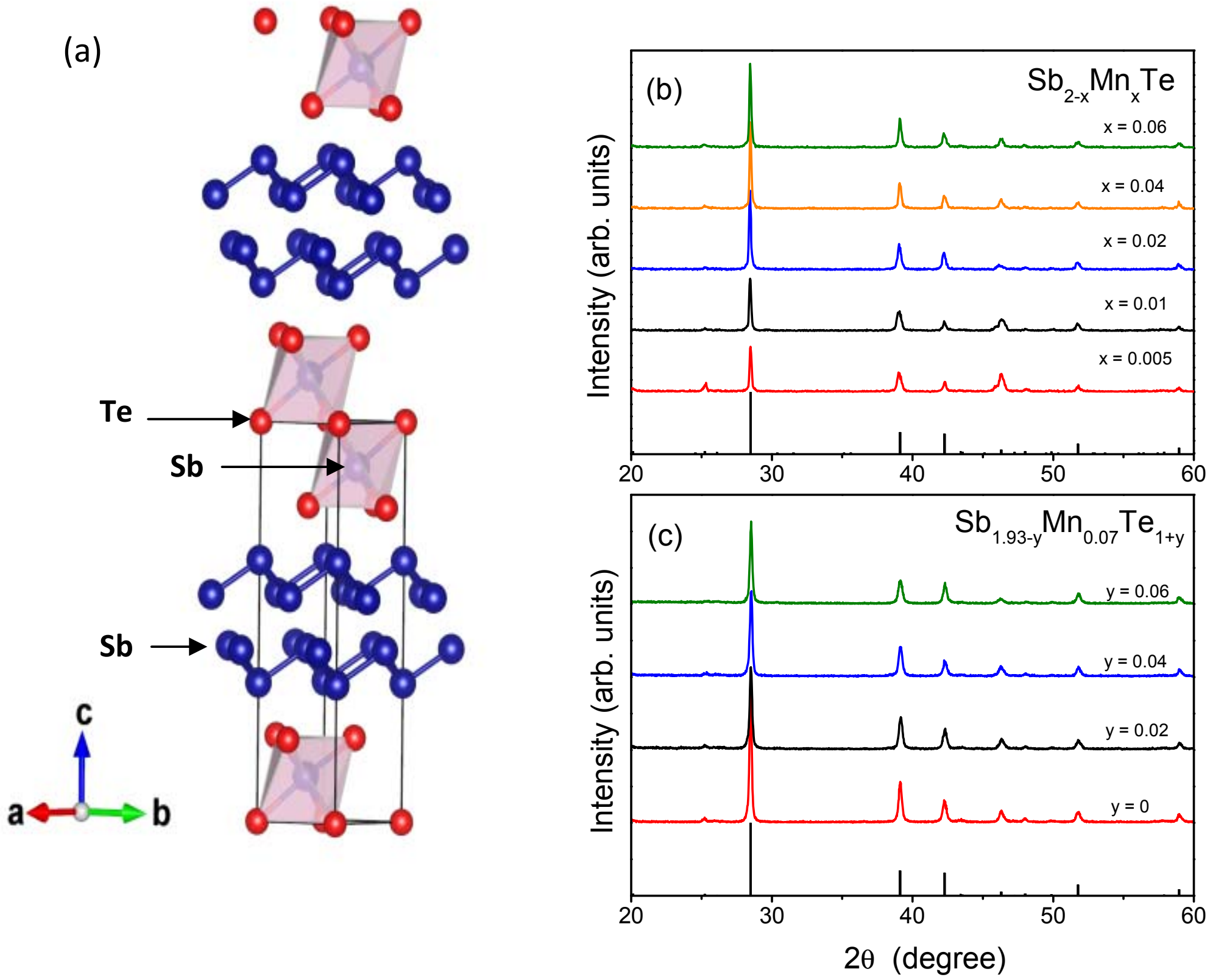

**Figure 2**

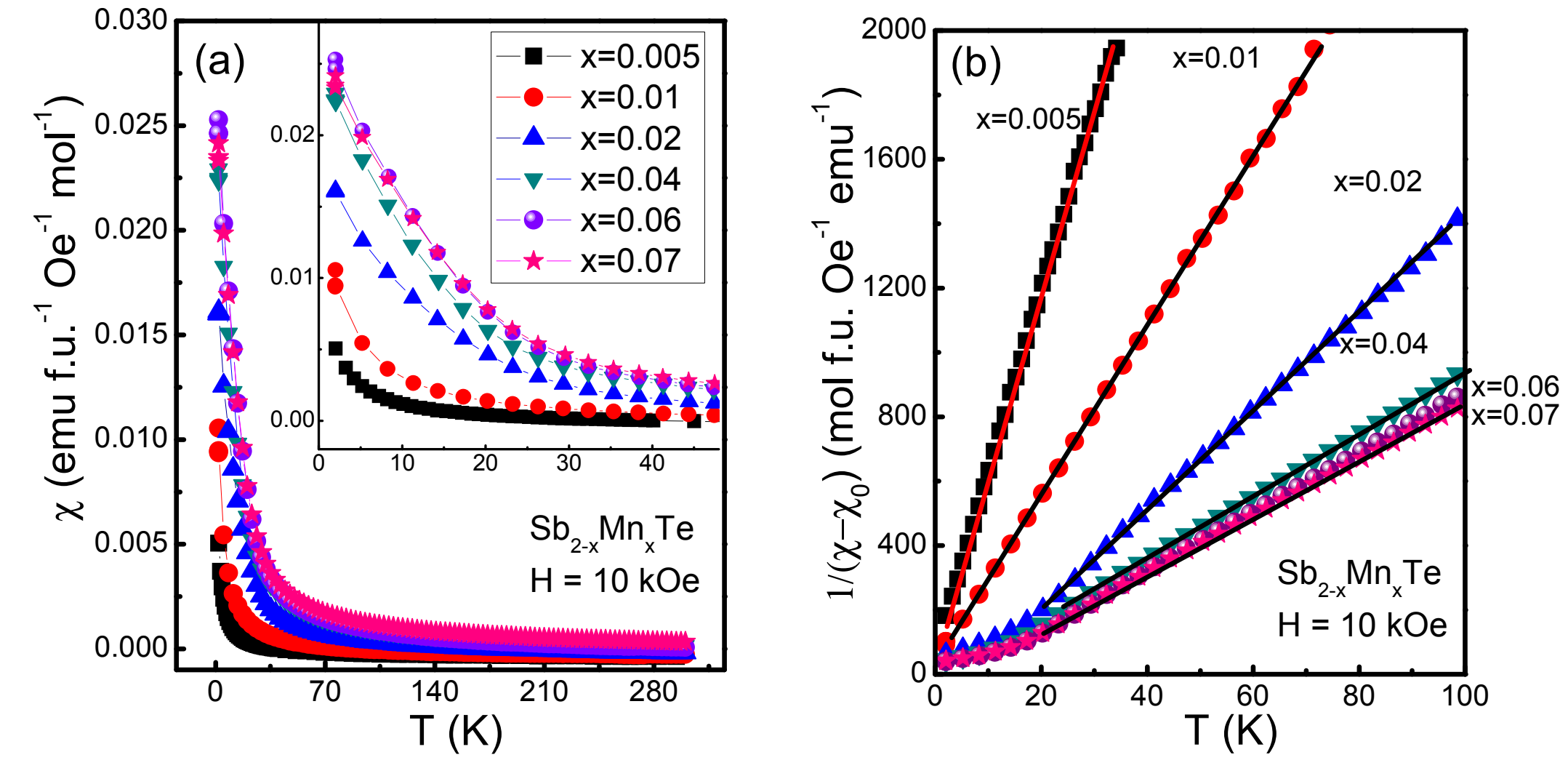

**Figure 3**

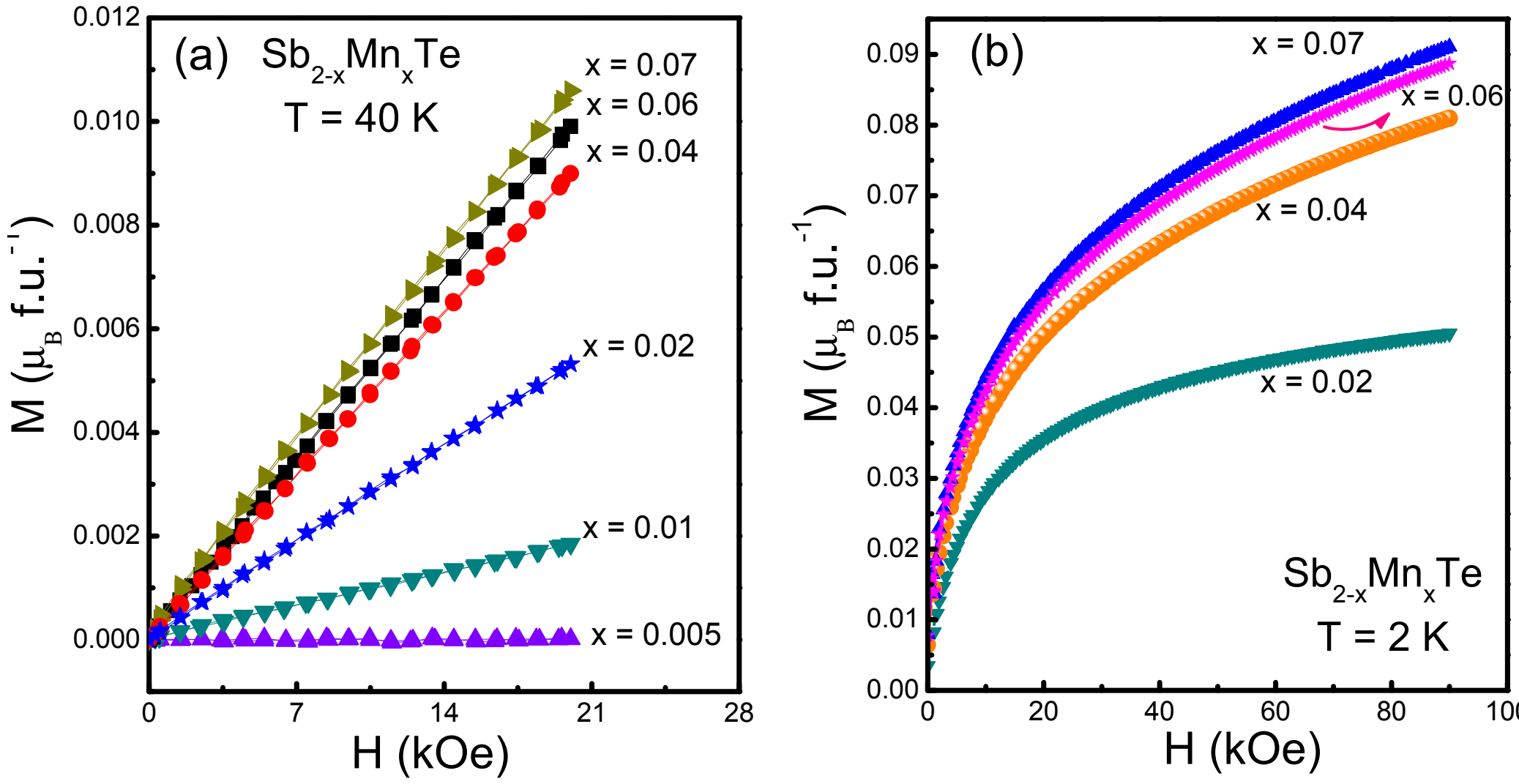

**Figure 4**

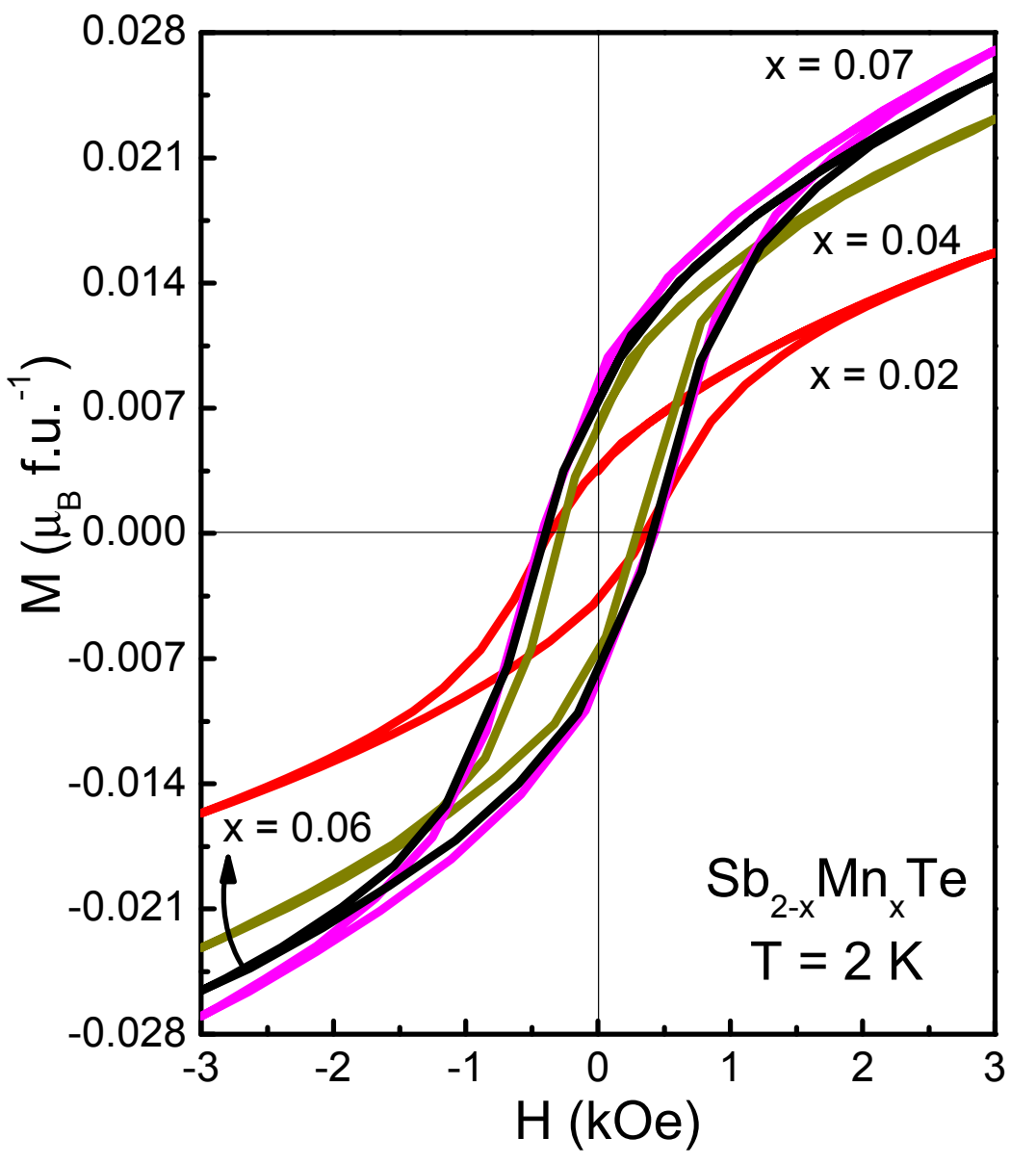

**Figure 5**

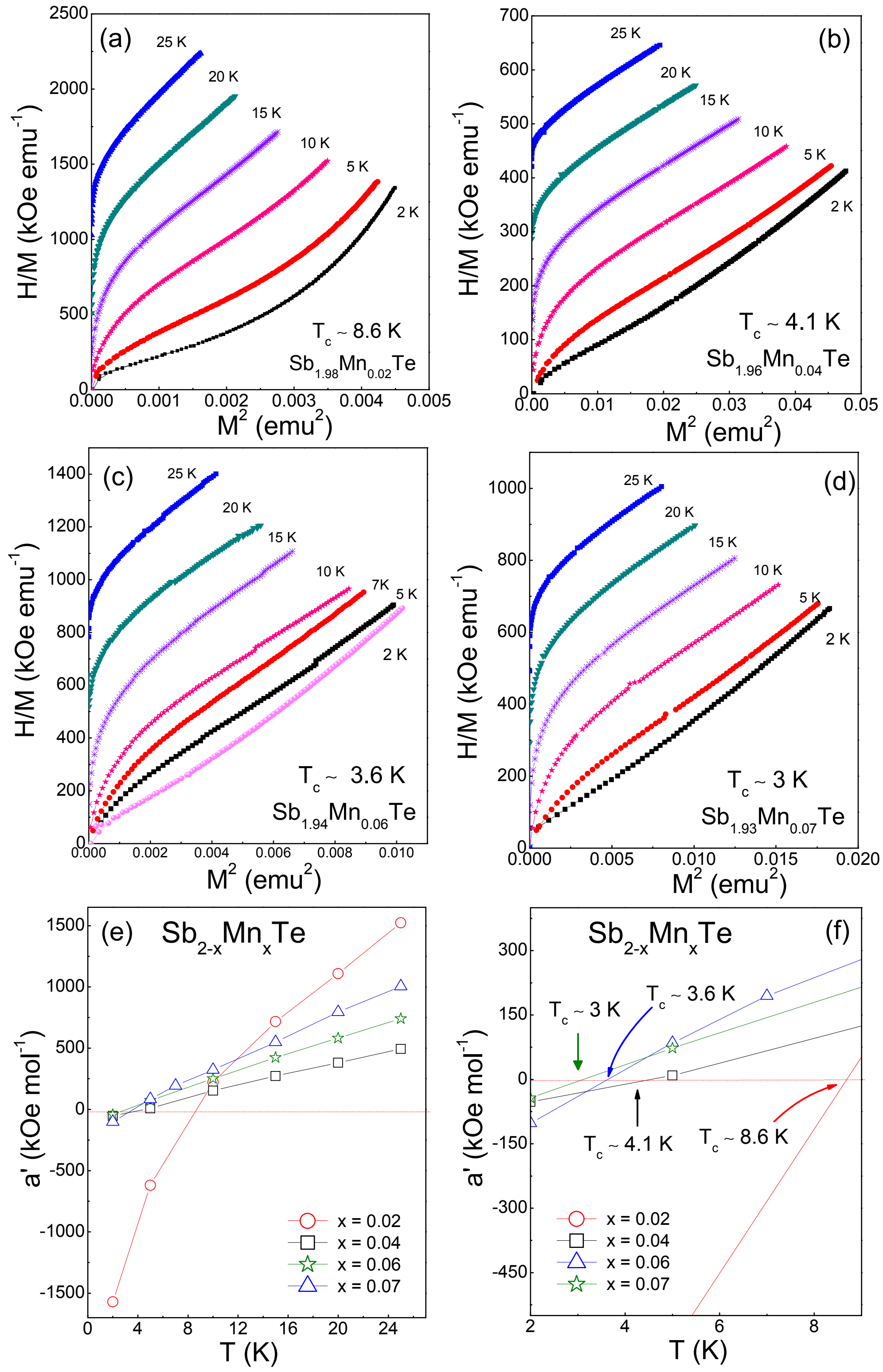

(a)
25 K
20 K
15 K
10 K
5 K
2 K
T$_c$ ~ 8.6 K
$Sb_{1.98}Mn_{0.02}Te$
H/M (kOe emu$^{-1}$)
M$^2$ (emu$^2$)
(b)
T$_c$ ~ 4.1 K
$Sb_{1.96}Mn_{0.04}Te$
(c)
7K
T$_c$ ~ 3.6 K
$Sb_{1.94}Mn_{0.06}Te$
(d)
T$_c$ ~ 3 K
$Sb_{1.93}Mn_{0.07}Te$
(e)
$Sb_{2-x}Mn_xTe$
a' (kOe mol$^{-1}$)
T (K)
x = 0.02
x = 0.04
x = 0.06
x = 0.07
(f)
T$_c$ ~ 3 K
T$_c$ ~ 3.6 K
T$_c$ ~ 4.1 K
T$_c$ ~ 8.6 K

**Figure 6**

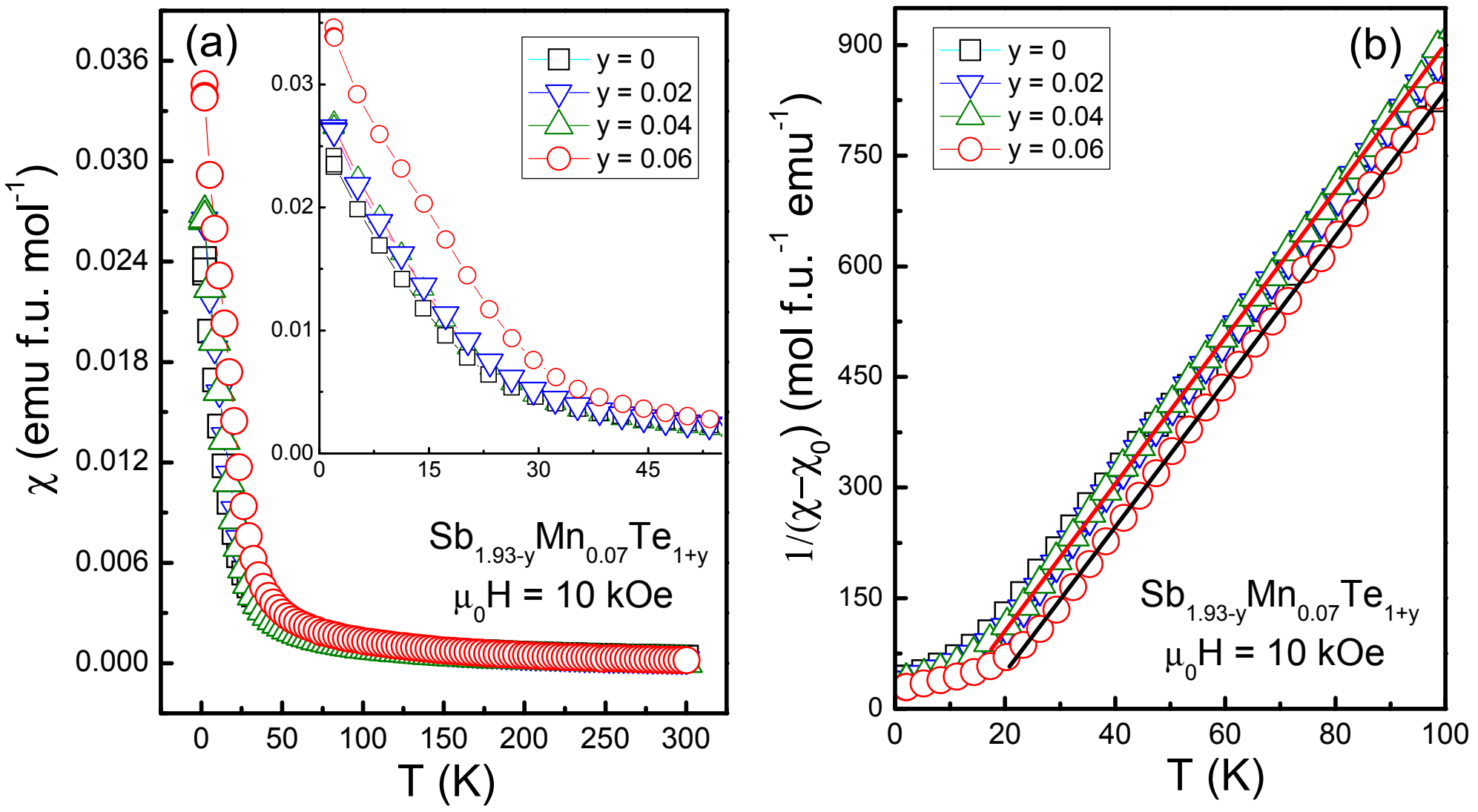

**Figure 7**

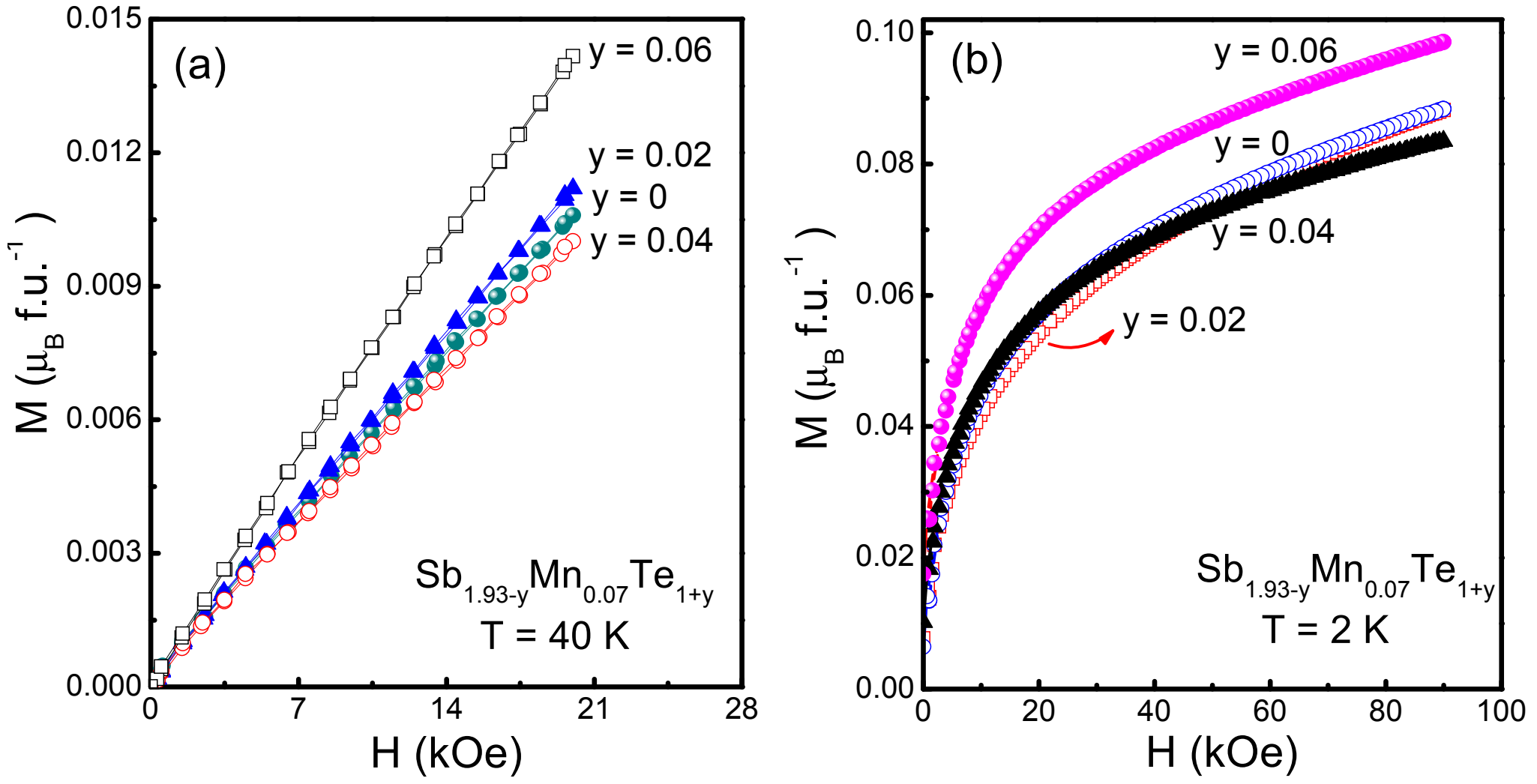

**Figure 8**

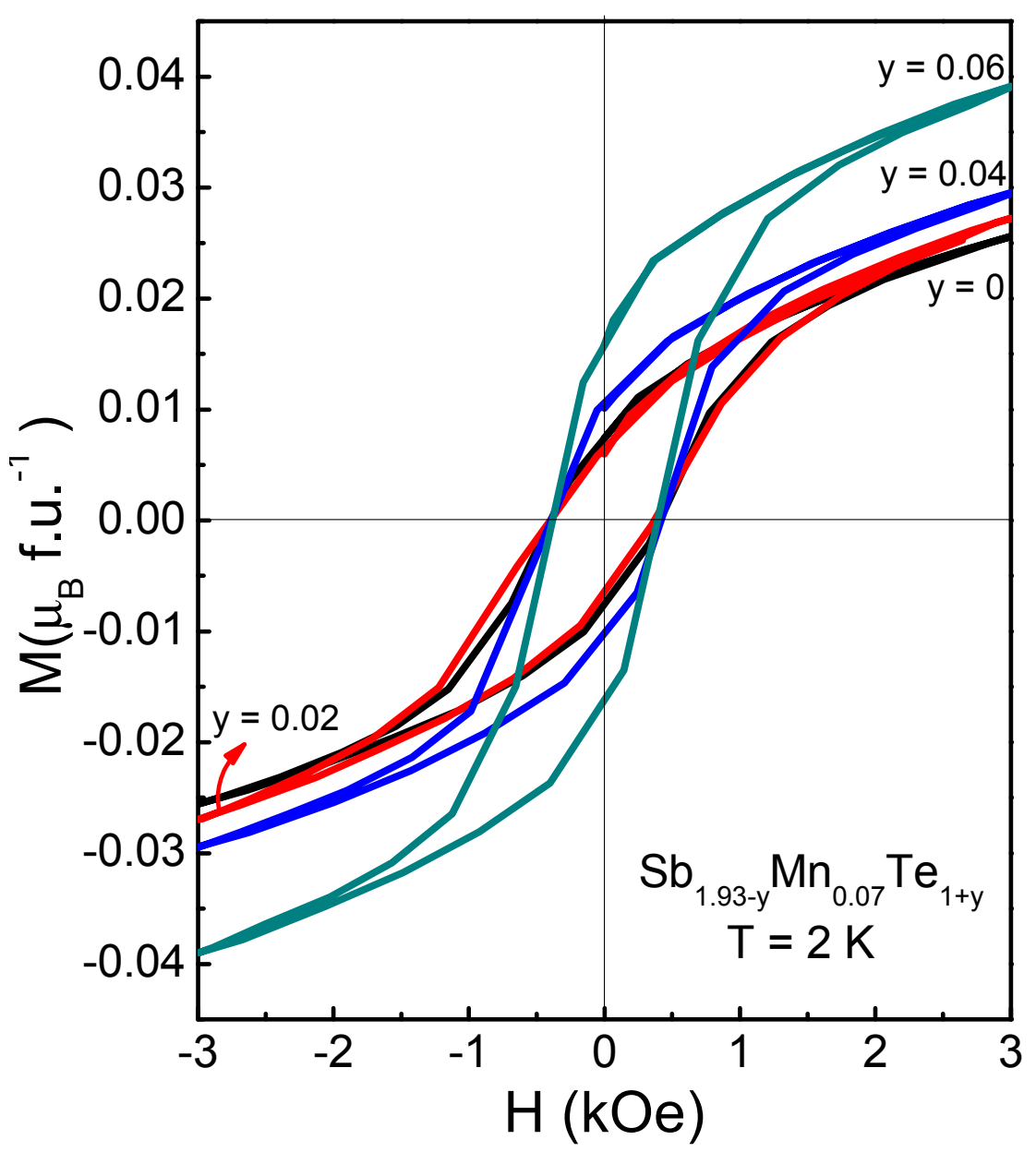

**Figure 9**

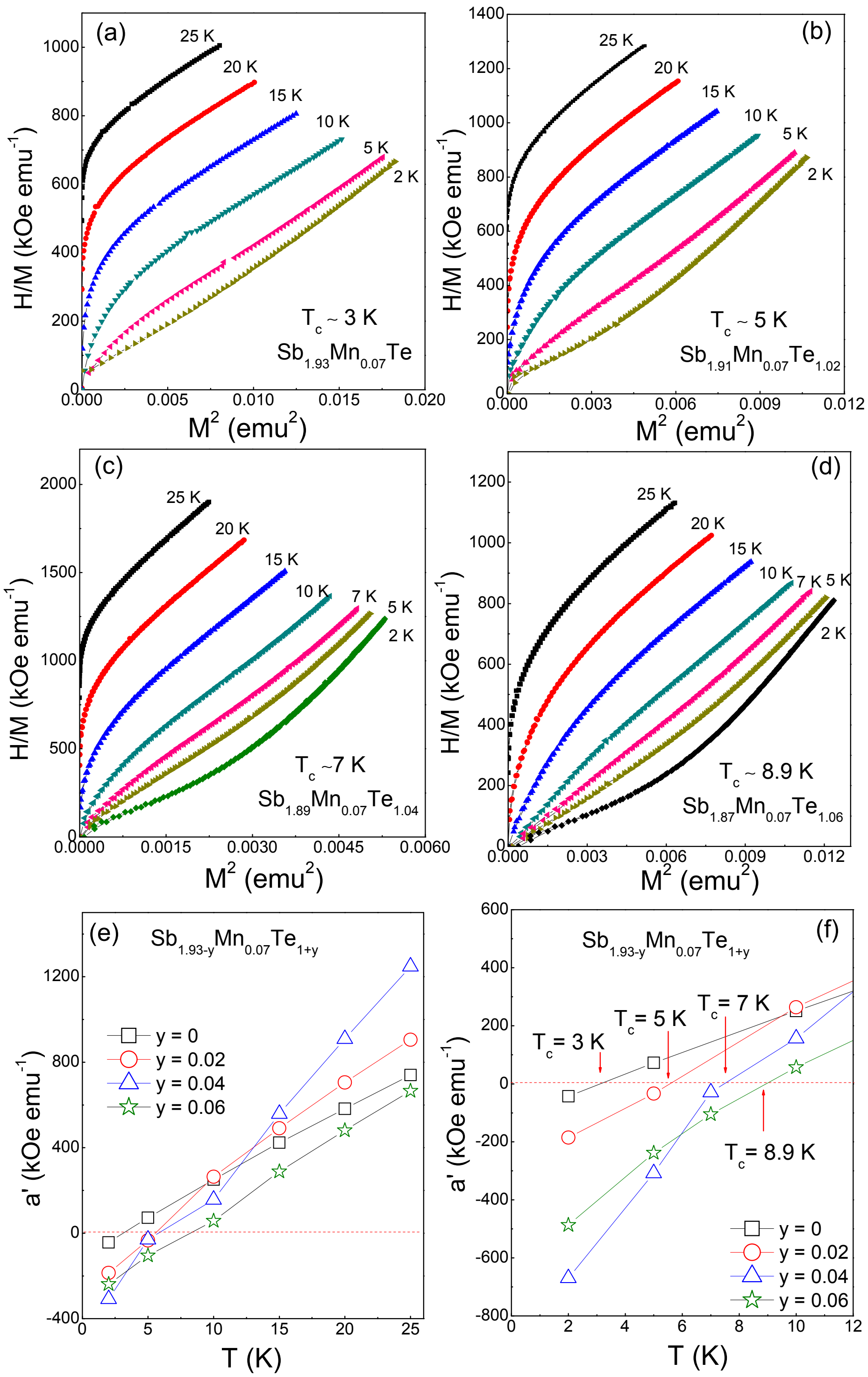

(a)
25 K
20 K
15 K
10 K
5 K
2 K
T_c ~ 3 K
Sb1.93Mn0.07Te
H/M (kOe emu^-1)
M^2 (emu^2)
(b)
T_c ~ 5 K
Sb1.91Mn0.07Te1.02
(c)
7 K
T_c ~7 K
Sb1.89Mn0.07Te1.04
(d)
T_c ~ 8.9 K
Sb1.87Mn0.07Te1.06
(e)
Sb1.93-yMn0.07Te1+y
y = 0
y = 0.02
y = 0.04
y = 0.06
a' (kOe emu^-1)
T (K)
(f)
T_c = 3 K
T_c = 5 K
T_c = 7 K
T_c = 8.9 K